\documentclass[12pt]{article}
\usepackage{amsmath}
\usepackage{amsfonts}
\usepackage{graphicx}

\begin{document}

\begin{center}
\textbf{\large Light trapping and guidance in plasmonic nanocrystals}\\*[%
0.1cm]
\textbf{Maxim Sukharev and Tamar Seideman}\\*[0.2cm]

\textit{Department of Chemistry, Northwestern University,\\[0pt]
2145 Sheridan Road, Evanston, IL. 60208-3113 USA}\\[0pt]

\textbf{ABSTRACT}
\end{center}

\vspace*{-0.3cm}{\small \ We illustrate the possibility of light trapping
and funneling in periodic arrays of metallic nanoparticles. A controllable
minimum in the transmission spectra of such constructs arises from a
collective plasmon resonance phenomenon, where an incident plane wave
sharply localizes in the vertical direction, remaining delocalized in the
direction parallel to the crystal plane. Using hybrid arrays of different
structures or different materials, we apply the trapping effect to structure
the eigen-mode spectrum, introduce overlapping resonances, and hence direct
the light in space in a wavelength-sensitive fashion.}\\*[0.3cm]

\section{Introduction}

\label{Introduction}

Tremendous progress in fabrication methods of metallodielectric
nanostructures,\cite{Hutter2004} on the one hand, and rapid development of
laser technology,\cite{SlusherRevModPhys1999} on the other, have recently
enabled a wide variety of exciting opportunities for light manipulation in
subdiffraction length scales. These range from understanding of fundamental
concepts of plasmon-polariton dynamics,\cite{SPP-reviews} through coherent
and optimal control of electromagnetic energy propagation via metallic
structures,\cite{SPP-control} to different applications in modern optical
device technology,\cite{Atwater2001review} single atom-molecule
manipulations,\cite{VanDuyneScience2004} and ultrasensitive detection of
biological molecules.\cite{Alivisatos2004}

The physical basis of these applications is the surface plasmon resonance
phenomenon, owing to which nobel metal particles of nanosize strongly
enhance electromagnetic (EM) fields tuned to resonance with the frequency of
collective oscillation of conductive electrons confined in the nanoparticle.
It follows from the Mie theory,\cite{Kreibig1995} (and was seen numerically
as well as experimentally) that the resonance wavelength depends crucially
on the particle size and shape. This sensitivity is the origin of many of
the applications of nobel metal nanoparticles, including their use as
sensors,\cite{Rindzevicius2005,Lu2006} medical diagnostics,\cite%
{El-Sayed2005} and standards.\cite{Reinhard2005}

Like the response to incident light of single nanoparticles (NPs), the
collective response of NP chains has been the topic of significant
experimental\cite{NP-chains-experiments} and theoretical interest in recent
years.\cite{NP-chains-theory} Particularly interesting is the possibility of
guiding light in the nanoscale via metallic NP chains and junctions.\cite%
{Bowden1989} Here, one excites the construct locally in space at one end of
the chain, typically by means of a nanosized tip, and detects the
transmitted energy, again with spatial resolution, at the other end, e.g.,
by means of a dye molecule. Much progress on this question not withstanding,
the adverse effect of losses have been identified as the major hurdle to
realizing waveguides based on NP chains.\cite{SPP-reviews} Films\cite%
{Films-Refs} and wires\cite{Wires-Refs} suffer much less from losses, but
lack the local enhancement and mode specificity of NPs.

Similarly fascinating, are the properties and applications of plasmonic
nanocrystals. By nanocrystals, we refer to ordered two- or three-dimensional
arrays of nanoparticles, periodic in the lateral direction, that are
collectively excited by an incident (delocalized) plane wave.
Metallodielectric crystals exhibit omnidirectional band gaps\cite%
{JoannopoulosPRB1996} along with a broadband absorption.\cite{Fan2005}
Experimental measurements of optical properties of \ one-\cite%
{GiessenPRL2003} and two-dimensional\cite{FriendNanoLett2006} periodic metal
nanostructures have been successfully performed. We also note that, not only
periodic structures composed of NPs exhibit interesting phenomena but also
periodic arrays of sub-wavelength holes have surprising optical transmission.%
\cite{LezecNature1998} The physics underlying such constructs has aspects in
common with that underlying finite NP chains and junctions, it relies on the
collective response of ordered NPs that are sufficiently closely-spaced to
interact via dipole and quadruple forces. The new features arise from the
periodicity in one direction, which gives rise to certain solid state
features, such as a band structure, and justifies the name \textquotedblleft
crystals\textquotedblright . The practical interest in nanoplasmonic
crystals steams from the premise of scaling to the nano-domain several of
the attractive properties of micron-sized analogs, such as photonic crystals
and band-gap materials, which have been extensively studied experimentally
and theoretically for their many applications.\cite{Soukoulis-review2002} Of
particular interest are properties such as light-induced transparency,\cite%
{He2006} self-collimation,\cite{Lu2006} and transmission through sharp bends.%
\cite{Mekis1996} While the essential physics responsible for these features
in the macro-world is quite different from that underlying the nanoscale
variant, the effects, as illustrated below, are similar.

We show in the following sections that plasmonic nanocrystals offer also new
and fundamentally interesting phenomena, that are unique to the nanosize and
do not have a direct analog in their micron-sized counterparts. In
particular, we illustrate the possibilities of introducing overlapping
resonances in the eigen-mode spectrum as a control tool, and of funneling
the incident plane wave in a predetermined direction with wavelength
selectivity. In the next section we briefly outline the theory and the
numerical approach taken. Section \ref{Results} presents and discusses our
results and the final section concludes, with an outlook to future research.

\section{Theory}

\label{Theory}

The interaction between the light and the metal nanoconstructs is simulated
using a finite-difference time-domain approach.\cite{Taflove2000} We
restrict attention to two-dimensional structures, and consider the TE$_{z}$
mode (transverse-electric mode with respect to $z$) of the EM field.
Maxwell's equations in this case read, 
\begin{gather}
\varepsilon _{eff}\frac{\partial E_{x}}{\partial t}=\frac{\partial H_{z}}{%
\partial y}-J_{x},  \notag \\
\varepsilon _{eff}\frac{\partial E_{y}}{\partial t}=-\frac{\partial H_{z}}{%
\partial x}-J_{y},  \label{Maxwell equations} \\
\mu _{0}\frac{\partial H_{z}}{\partial t}=\frac{\partial E_{x}}{\partial y}-%
\frac{\partial E_{y}}{\partial x},  \notag
\end{gather}%
where $E_{x}$, $E_{y}$, and $H_{z}$ denote the Cartesian components of the
EM field, $J_{x}$ and $J_{y}$ are the Cartesian components of the current
density, $\vec{J}$, $\varepsilon _{eff}$ is the effective dielectric
constant defined below, and $\mu _{0}$ is the magnetic permeability of free
space. The metal structures are described within the Drude model with a
complex valued, frequency-dependent dielectric constant,\cite{Huffman1983} 
\begin{equation}
\varepsilon \left( \omega \right) =\varepsilon _{0}\left( \varepsilon
_{\infty }-\frac{\omega _{p}^{2}}{\omega ^{2}+i\Gamma \omega }\right) ,
\label{Drude model}
\end{equation}%
where $\varepsilon _{0}$ is the electric permittivity of free space, $%
\varepsilon _{\infty }$ is the dimensionless infinite frequency limit of the
dielectric constant, $\omega _{p}$ is the bulk plasmon frequency, and $%
\Gamma $ is the damping rate. In the numerical simulations we use the
following two sets of parameters, which describe fairly well silver and gold
nanoparticles in the incident wavelength regime of interest. For silver
nanoconstructs, $\varepsilon _{\infty }=8.926$, $\omega _{p}=1.7601\times
10^{16}$ rad/sec, and $\Gamma =3.0841\times 10^{14}$ rad/sec. For gold
nanoconstructs, $\varepsilon _{\infty }=9.84$, $\omega _{p}=1.3673\times
10^{16}$ rad/sec, and $\Gamma =1.0179\times 10^{14}$ rad/sec.

In regions of space occupied by metallic particles, the material dispersion
gives rise to time-dependent current and Maxwell's equations (\ref{Maxwell
equations}) are supplemented by the additional set of equations for the
current density $\vec{J}$,\cite{Ziolkowski1995,GrayPRB2003} 
\begin{equation}
\frac{\partial \vec{J}}{\partial t}=a\vec{J}+b\vec{E},  \label{current}
\end{equation}%
where $a=-\Gamma $, $b=\varepsilon _{0}\omega _{p}^{2}$ and the effective
dielectric constant in Eq. (\ref{Maxwell equations}) is $\varepsilon
_{eff}=\varepsilon _{0}\varepsilon _{\infty }$ . In the surrounding free
space $\varepsilon _{eff}=\varepsilon _{0}$ and $a=b=0$. In order to avoid
nonphysical reflections of outgoing electromagnetic waves from the grid
boundaries, we implement perfectly matched layers (PML) of absorbing
boundaries\cite{BerengerPML} of depth of $16$ spatial steps each.

As an initial condition we employ a plane wave generated along a line, each
point of which is driven by a time-dependent function of the form, 
\begin{equation}
\vec{E}_{\text{inc}}\left( t\right) =\vec{e}_{x}E\left( t\right) \cos \omega
t,  \label{incident field}
\end{equation}%
where $\vec{e}_{x}$ is a unit vector along the $x$-axis, $E\left( t\right)
=E_{0}\sin ^{2}\left( \pi t/\tau \right) $ is the pulse envelope, $\tau $ is
the pulse duration ($E\left( t>\tau \right) =0$), and $\omega $ is the
optical frequency. In all simulations $\tau =50$ fs, the total propagation
time is $100$\ fs, with a spatial step size of $\delta x=\delta y=1.5$
\thinspace nm and a temporal step size of $\delta t=\delta x/\left(
1.5c\right) $, where $c$ denotes the speed of light in vacuum. All
simulations have been performed on distributed memory parallel computers at
the National Energy Research Scientific Computing Center and San Diego
Supercomputer Center. The parallel technique used in our simulations is
described in detail in Ref. 31.

\section{Results and Discussion}

\label{Results}

We open this section by exploring the properties of the simplest plasmonic
crystal that could be envisioned. This construct will serve to introduce the
basic phenomena that underlie several of the much more complicated
constructs discussed below. The nano-crystal envisioned is schematically
depicted in the inset of Fig. 1. It is composed of a periodic array of
silver NPs with $55$ nm diameter and $66$ nm center-to-center distance. The
NPs are excited by an $x$-polarized plane wave propagating in the negative $%
y $-direction, and the EM energy is detected along the horizontal line shown
in dashed in the inset. The transmission through the nanostructures is
expressed in terms of the ratio of the time-averaged EM energy in the
presence of the periodic array, $W_{\text{total}}$, to that in its absence, $%
W_{\text{inc}}$. This ratio is shown in the main frame of Fig. 1 for five
arrays with different number of layers as a function of the incident
wavelength. The transmission is close to perfect throughout the wavelength
regime of relevance, safe for a narrow window centered at $\lambda \simeq
360 $ nm, where it drops to a sharp minimum. The transmission minimum is
sharper the larger the number of layers, converging to essentially zero
transmission ($W_{\text{total}}/W_{\text{inc}}\approx 2\times 10^{-7}$) with 
$5$ layers.

The origin of the transmission minimum is the excitation of longitudinal ($x$%
-polarized) plasmons, due to which an incident light propagating along the $%
y $-axis bends as it enters the periodic array and is guided along the
horizontal axis. The efficiency with which the light takes the sharp corner,
and the spatial localization of its subsequent path are investigated below.
Our results provide a likely explanation to the finding of Ref. 19, where a
strong broadband absorption in metallic photonic crystals is reported. As
shown below, the resonance feature is due to an interesting trapping
phenomenon, whereby the electromagnetic energy is localized along the
surface of the crystal. Such collective waves that are fully delocalized
laterally but sharply localized in the vertical direction are familiar in
electronic systems, where they arise from image potentials.\cite%
{image-potentials} We are not aware of previous studies that have observed
such phenomena in plasmonic waves.

Both the broad transmission window and the ability of the crystal to
re-direct the propagating EM energy by $90$ degrees and guide it in parallel
to the crystal plane at the transmission minimum wavelength are of
significant practical interest, as they address two long researched goals of
nanophotonics.\cite{SPP-reviews} More interestingly from a fundamental
perspective, the resonance phenomenon that underlies the effect offers
exciting opportunities for optical control. Such resonance features have
been intensively studied in atomic and molecular physics, where their
manifestation in radiative lineshapes is conveniently described within the
Feschbach partitioning of space into a bound and a scattering manifold. The
present system can be analyzed using the machinery of Fano and Feschbach by
analogy to the well-studied case of matter waves, but offers the advantage
over the latter case that the interaction coupling the discrete state to the
continuum can be systematically tuned by design of the array.

We proceed by exploring the dependence of the resonance feature on the
experimentally variable parameters of the array. In Fig. 2 we focus on the
single layer array and gradually modify the resonance properties by coating
the silver nano-spheres by a non-dispersive insulator. The inset of Fig. 2
depicts schematically the array envisioned. We remark that core-shell
nanoparticles of the sort shown in the inset of Fig. 2 have been synthesized
in several recent studies that have noted the potential applications and
interesting properties of such constructs. Several trends in Fig. 2, and in
related studies that are not shown here, can be qualitatively understood
from the Mie expression for the extinction cross section of a single sphere.
Both an increase in the dielectric constant of the core material and an
increase in the dielectric constant of the shell material red-shift the
resonance position. An increase in the shell dielectric constant has the
additional effect of broadening the resonance, see Fig. 2.

Of particular interest are devices that support multiple resonances. In
molecular systems, the case of overlapping resonances has been studied
extensively for its fundamental interest as well as for intriguing
opportunities for coherent control.\cite{OverlapRes} In the nanoplasmonics
considered here, both the case of isolated and that of overlapping multiple
resonances are of fundamental and practical interest, and both can be
readily realized by proper design of the nano-crystal. Figure 3 explores
both the case of non-overlapping and that of overlapping resonances. To that
end we consider hybrid nanocrystals consisting of ordered layers of two
types of metallic nanoparticles and include both spheres and ellipsoids. The
insets depict the two nano-crystals considered, one of which (inset a)
combines silver and gold nano-spheres and the other (inset b) considers
alternating layers of spheres and ellipsoids, the top two of which are
silver, and the bottom two gold. The combination of Au and Ag NPs provides a
model for study of non-overlapping resonances, since the plasmon resonance
wavelengths of Ag and Au are separated by ca $120$ nm. The combination of
spheres and ellipsoids provides a model for study of overlapping resonances,
since the shape-dependence of the resonance wavelength is small, leading to
a shift of the line center comparable to the resonance width.

The main frame of Fig. 3 shows simulations of the ratio, $W_{\text{total}%
}/W_{\text{inc}}$, for the two hybrid constructs depicted in the insets,
with the solid curve corresponding to the bimetallic nano-crystal of inset
(a) and the dashed curve corresponding to the mixed crystal of inset (b).
Both spectra show the expected well-defined double-peak structure. The
spectrum generated by construct (b) exhibits also significant broadening and
additional structure, arising from the partially overlapping resonances of
the spheres and ellipsoids.

The results of Fig. 3 can be understood through an eigenmode analysis of the
elements constituting the nanocrystals. The eigenmodes of a given construct
are computed by excitation with a broad band, short ($0.36$ fs) pulse,
followed by long time ($2$ ps) simulation of the $x$-component of the
resultant scattered electric field. Fourier transform of the field yields
the power spectrum $I(\lambda )$. The solid and dashed curves in Fig. 4
correspond to the cases of single (periodic) layers of silver and gold
nano-spheres, respectively, whereas the dot-dashed curve corresponds to one
Ag and one Au nano-spheres layer and the dotted curve to one Ag spheres and
one Ag ellipsoids layer. Several interesting features are noted, including
Fano-type interference between direct and resonant transmission resulting in
marked asymmetry (dashed curve), non-overlapping resonances (dot-dashed
curve) and overlapping resonances (dotted curve).

The plasmon resonances of silver and gold linear periodic chains are
separated by almost $120$ nm, as illustrated in Fig. 4, resulting not only
in selective transmission of light by the hybrid nano-structure but also in
separation of different resonant wavelengths in space. For example,
wavelengths near the plasmon resonance of silver chains, $\lambda \simeq 380$
nm, are trapped in the first two silver layers (see the insets in Fig. 3),
whereas wavelengths near the plasmon resonance of gold chains ($\lambda
\simeq 500$ nm) are transmitted through the silver layers and excite the
bottom two gold chains. The wavelength-dependent transparency of different
parts of a nano-structure is due to the fact that the effective thickness of
the skin-layer at the wavelengths considered is of the same order of
magnitude as the size of the NPs. The skin-depth ($d$) at optical
frequencies ($\omega $) much larger than the plasmon frequency ($\omega _{p}$%
) can be estimated\cite{Greiner1998} as $d\simeq c/\omega _{p}$, which leads
to $17$ nm and $22$ nm for silver and gold, respectively.

The wavelength-sensitive transparency of different parts of nanostructures
suggests exciting opportunities for use of non-symmetric (with respect to $y$%
) constructs for light manipulation and directed transport. Two examples of
light guidance are presented in Fig. 5, where we simulate hybrid structures
composed of silver and gold NPs that are finite in both $x$- and $y$%
-directions. At $\lambda =340$ nm, for instance, the left part of the
nano-crystal depicted in the inset (a) of Fig. 5 traps the light, while the
right part, composed of gold NPs is transparent. Excited at their plasmon
resonance wavelength, silver layers guide the EM energy to the right. The
structure shown in inset (b) of Fig. 5, supports both horizontal directions
for EM energy guiding. This is illustrated in Fig. 6, where we plot the
spatial time-averaged distribution of the ratio, $W_{\text{total}}/W_{\text{%
inc}}$, at the incident wavelength corresponding to the plasmon resonance of
silver NPs. Clearly, an incident plane wave, initially symmetric, is broken
by the hybrid structure into two parts, one of which is guided to the left,
and the other to the right.

In summary, we illustrated selective trapping and guidance of light by
periodic metallic nanocrystals. Owing to the excitation of collective
longitudinal plasmons in linear chains of closely spaced NPs, an incident
wave can be selectively localized in the direction vertical to the light
propagation, remaining delocalized laterally. By proper device design, EM
excitation can be thus efficiently guided in space. A variety of extensions
of the phenomena of selective light trapping and funneling can be
envisioned. Three dimensional nanocrystals could be used to guide EM energy
in 3D space. The interaction of white light with periodic arrays could serve
to make interesting optical sources. Likewise inviting is the possibility of
guiding light in circular or helical pathways using properly structured
plasmonic nanoconstructs.

Acknowledgement.\textbf{\ }This work was supported in part by the NCLT
program of the National Science Foundation (ESI-0426328) at the Material
Research Institute of Northwestern University. We acknowledge the National
Energy Research Scientific Computing Center, supported by the Office of
Science of the U.S. Department of Energy under Contract No.
DE-AC03-76SF00098, and the San Diego Supercomputer Center under Grant No.
PHY050001, for computational resources. We thank Joseph Yelk for
computational assistance.

\newpage

\begin{center}
\bigskip {\Large Figures}
\end{center}

\begin{figure}[htbp]
\centering\includegraphics[width=\linewidth]{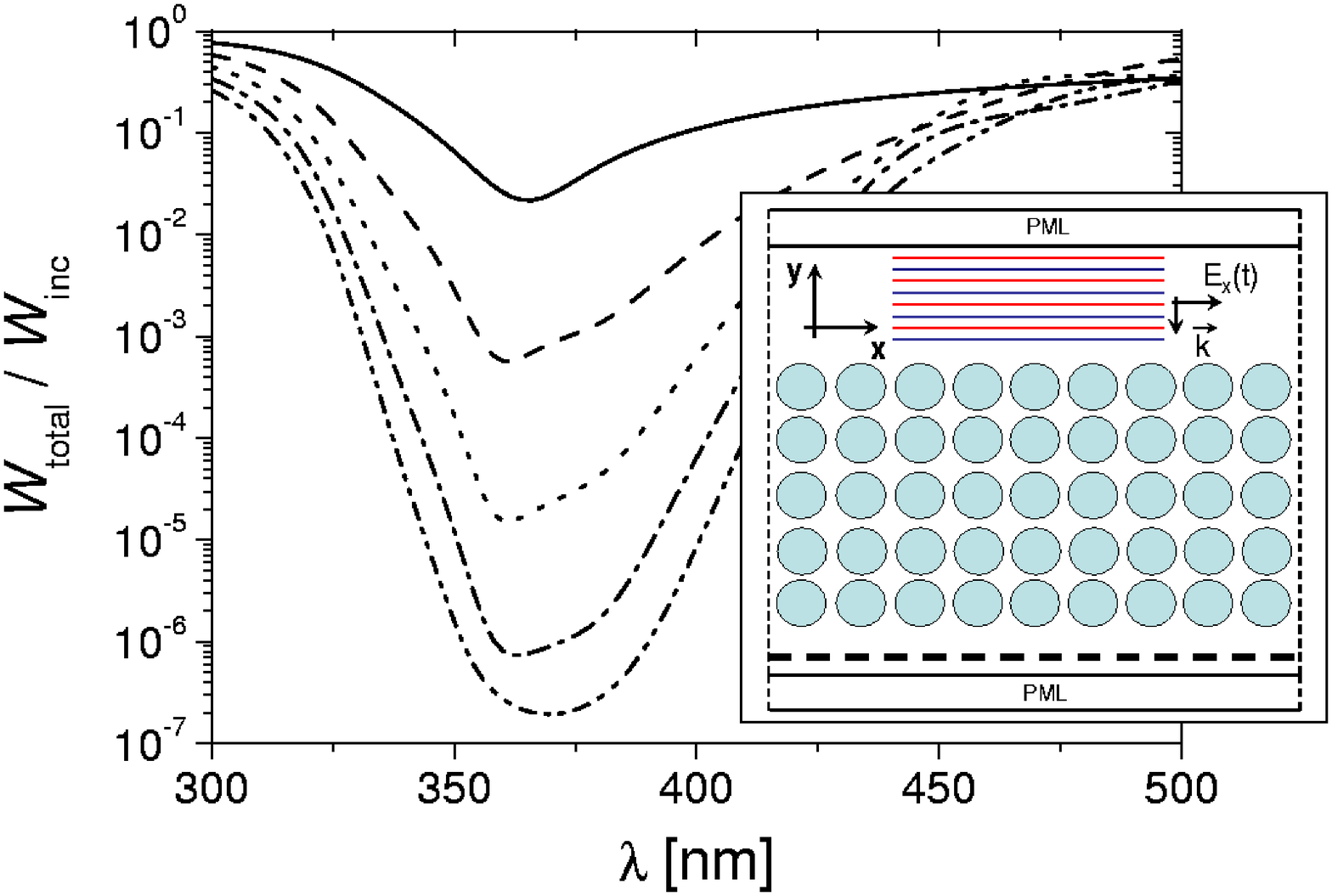}
\caption{\label{fig1}
The ratio of time-averaged EM energy calculated in the presence and
in the absence of the nanoconstruct schematically depicted in the inset as a
function of the incident wavelength, $\protect\lambda $. The solid curve
shows the ratio for a single horizontal layer of NPs, the dashed curve for
two layers, the dotted curve for three layers, the dash-dotted curve four
layers, and the dash-dot-dotted curve for five layers. The inset shows the
geometry envisioned, where the dashed vertical lines represent periodic
boundaries in the lateral direction. The nanoconstruct is excited by an $x$%
-polarized plane wave propagating in the negative $y$ direction and the EM
energy is calculated along the horizontal line shown as a dashed line
beneath the structure in the far-field zone (the distance of the detection
line from the crystal is $600$ nm).}
\end{figure}

\newpage

\begin{figure}[htbp]
\centering\includegraphics[width=\linewidth]{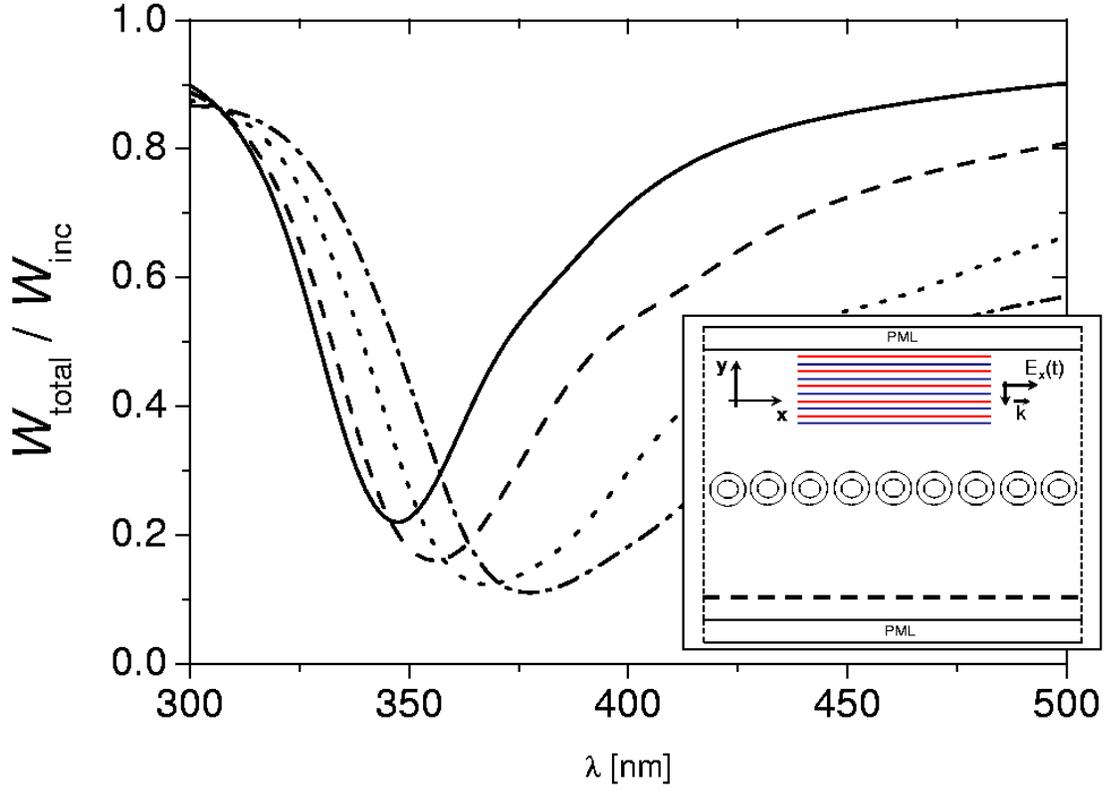}
\caption{\label{fig2}
As in Fig. 1, for a single layer of silver NPs coated by an
insulator (see inset). The solid curve shows the ratio $W_{\text{total}}/W_{%
\text{inc}}$ for bare particles (no coating, corresponding to $\protect%
\varepsilon =1$), the dashed curve for coating dielectric constant of $%
\protect\varepsilon =1.5$, the dotted curve for $\protect\varepsilon =2.5$,
and the dash-dotted curve for $\protect\varepsilon =3.5$. The silver NP
diameter is $35$\ nm, the diameter of the dielectric coating layer is $55$\
nm, and the center-to-center distance is $66$\ nm.}
\end{figure}

\newpage

\begin{figure}[htbp]
\centering\includegraphics[width=\linewidth]{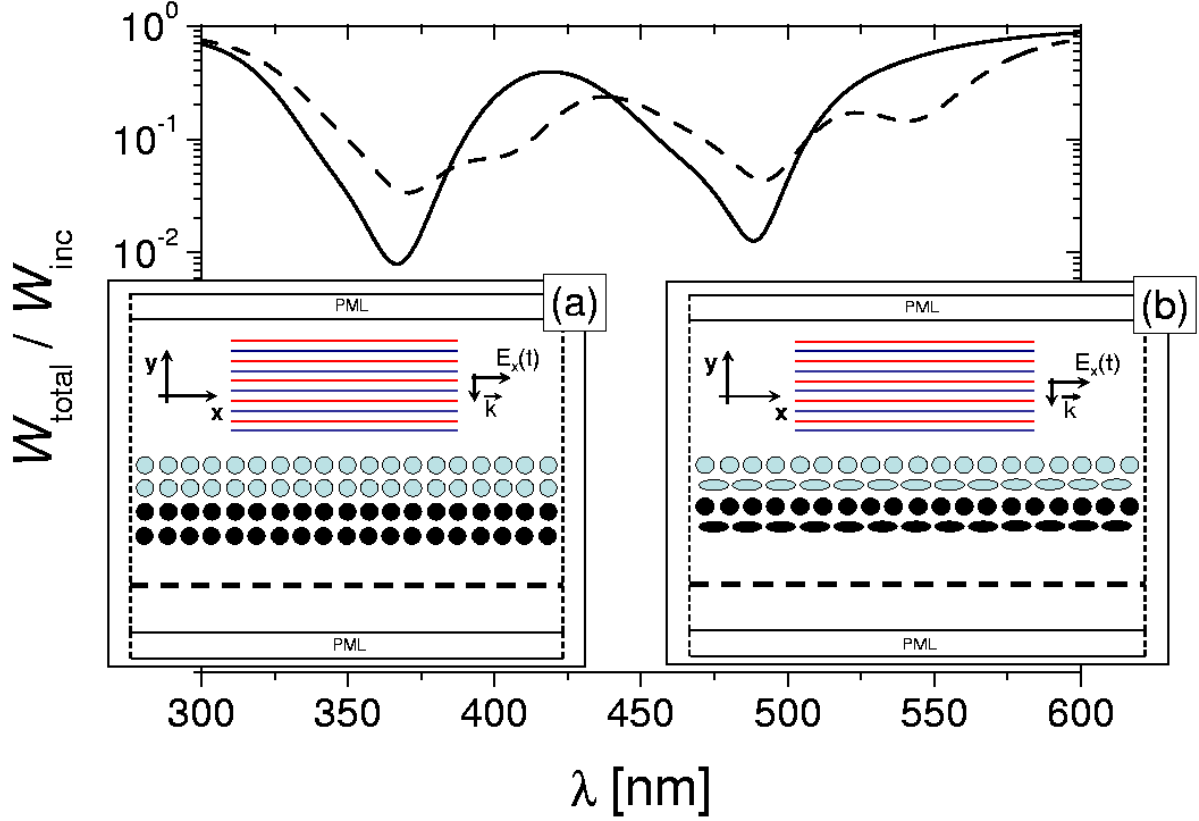}
\caption{\label{fig3}
The ratio $W_{\text{total}}/W_{\text{inc}}$ as a function of the
incident wavelength for the two nanoconstructs shown schematically in the
insets. (a) A silver-gold hybrid structure composed of spheres of diameter $%
32$\ nm and center-to-center distance $42$\ nm. (b) A silver-gold hybrid
structure composed of spheres and ellipsoids with major (minor) axes of $20$
($32$) \ nm and center-to-center distance of $42$ nm. Gold NPs are shown as
black, silver particles as grey spheres. The solid curve in the main frame
corresponds to structure (a) and the dashed curve to structure (b).}
\end{figure}

\newpage

\begin{figure}[htbp]
\centering\includegraphics[width=\linewidth]{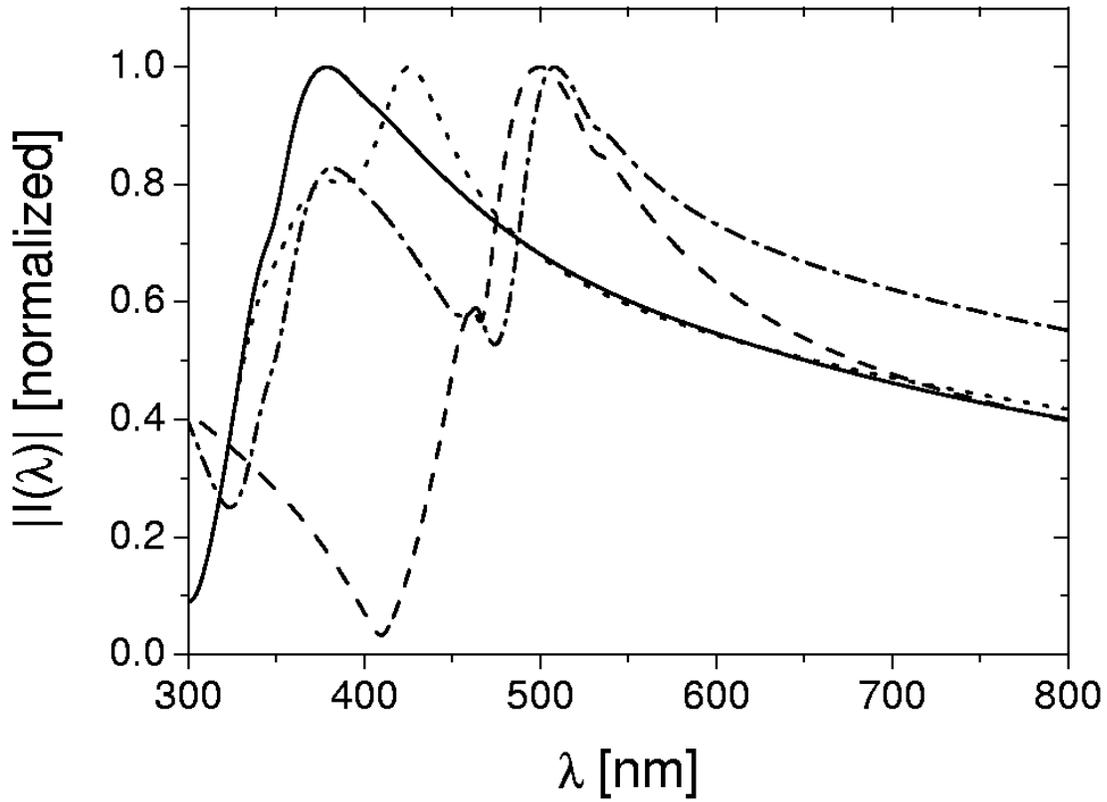}
\caption{\label{fig4}
Power spectrum, $\left\vert I\left( \protect\lambda \right)
\right\vert $,\ of the $x$-component, $E_{x}$, of the scattered electric
field in Fig. 3 as a function of the incident wavelength, $\protect\lambda $%
. The solid curve corresponds to a single layer of silver NPs, the dashed
curve to a layer of gold NPs, the dash-dotted curve to interacting layers of
silver and gold NPs, and the dotted curve to silver layers of spheres and
ellipsoids.}
\end{figure}

\newpage

\begin{figure}[htbp]
\centering\includegraphics[width=\linewidth]{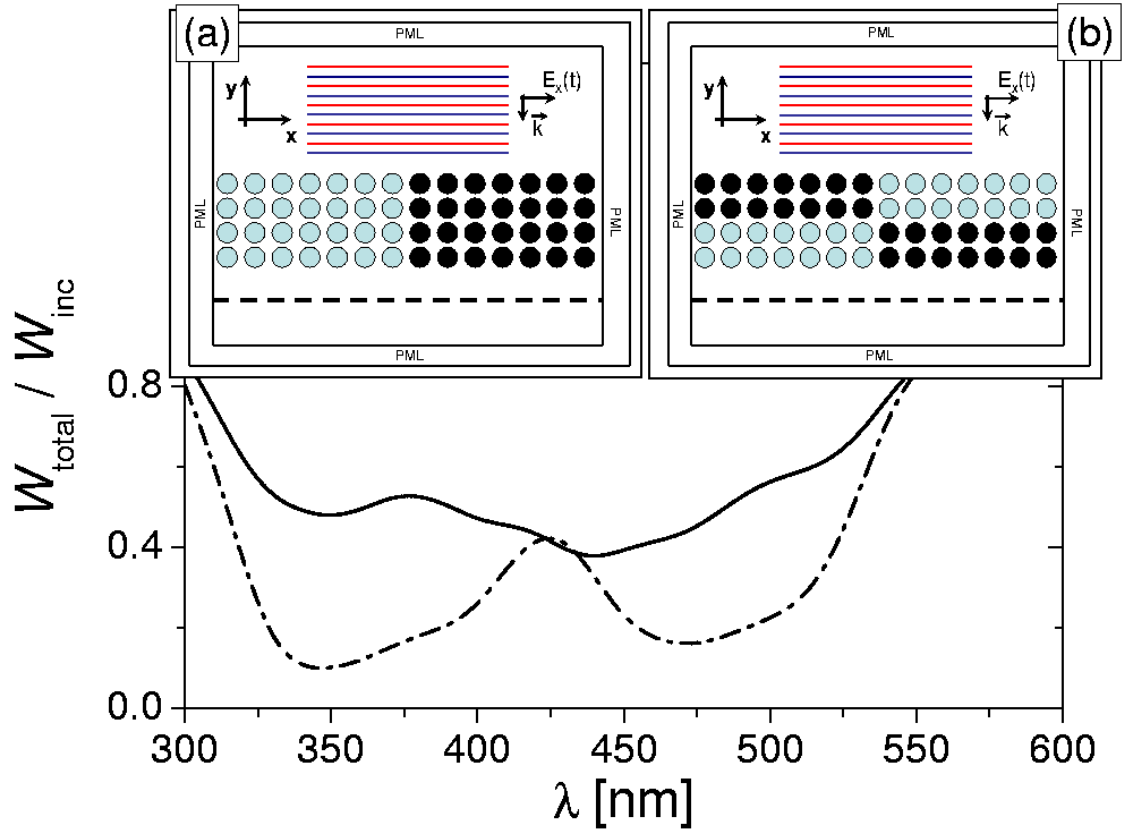}
\caption{\label{fig5}
The ratio $W_{\text{total}}/W_{\text{inc}}$ as a function of the
incident wavelength for the two finite nanoconstructs shown schematically in
the insets. Silver NPs are shown as grey spheres and gold NPs as black
spheres. The solid curve in the main frame presents simulations for
structure (a), and the dashed curve for structure (b). Both silver and gold
particles are $40$ nm in diameter and the center-to-center distance is $50$
nm.}
\end{figure}

\newpage

\begin{figure}[htbp]
\centering\includegraphics[width=\linewidth]{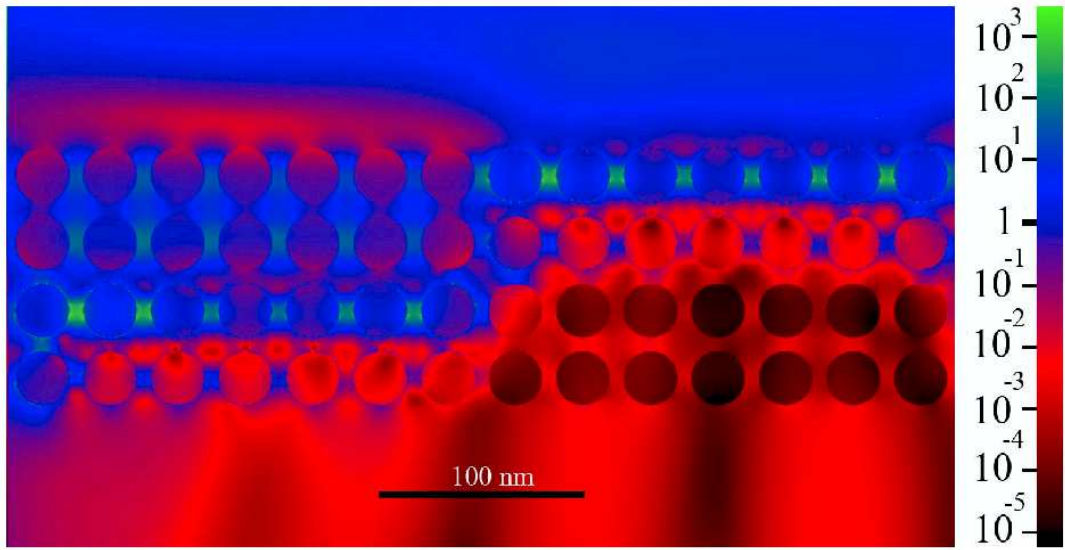}
\caption{\label{fig6}
Time-averaged ratio, $W_{\text{total}}/W_{\text{inc}}$, as a
function of $x$ and $y$ for the structure shown in inset (b) of Fig. 5 at an
incident wavelength of $\protect\lambda =340$ nm, corresponding to the
excitation of longitudinal plasmons in silver NPs.}
\end{figure}

\end{document}